\documentclass{emulateapj} 

\slugcomment{Submitted to ApJ}

\shorttitle{Low efficiency accretion in NGC~4565}
\shortauthors{Chiaberge et al.}

\begin{document}

\title{Low radiative efficiency accretion  at work in active galactic
nuclei: the nuclear spectral energy distribution of NGC~4565}


\author{M. Chiaberge\altaffilmark{1}}
\affil{Space Telescope Science Institute, 3700 San Martin Dr., Baltimore, MD 21218}
\email{chiab@stsci.edu}

\author{R. Gilli\altaffilmark{2}}
\affil{INAF - Osservatorio astronomico di Bologna, Via Ranzani 1, 40127 Bologna, Italy}


\author{F. D. Macchetto\altaffilmark{3}}
\affil{Space Telescope Science Institute, 3700 San Martin Dr., Baltimore, MD 21218}

\and 

\author{William B. Sparks}
\affil{Space Telescope Science Institute, 3700 San Martin Dr., Baltimore, MD 21218}


\altaffiltext{1}{On leave from INAF, Istituto di Radioastronomia, via P. Gobetti 101, I-40129 Bologna}
\altaffiltext{2}{Visiting Programmer, Space Telescope Science Institute}
\altaffiltext{3}{On assignment from ESA}


\begin{abstract}
We derive the spectral energy distribution (SED) of the nucleus of the
Seyfert galaxy \object{NGC~4565}. Despite its classification as a
Seyfert~2, the nuclear source is substantially unabsorbed. The
absorption we find from Chandra data ($N_H=2.5 \times 10^{21}$
cm$^{-2}$) is consistent with that produced by material in the
galactic disk of the host galaxy. HST images show a nuclear unresolved
source in all of the available observations, from the near-IR H band
to the optical U band. The SED is completely different from that of
Seyfert galaxies and QSO, as it appears basically ``flat'' in the
IR-optical region, with a small drop-off in the U-band. The location
of the object in diagnostic planes for low luminosity AGNs excludes a
jet origin for the optical nucleus, and its extremely low Eddington
ratio $L_o/L_{Edd}$ indicates that the radiation we observe is most
likely produced in a radiatively inefficient accretion flow (RIAF).
This would make NGC~4565 the first AGN in which an ADAF-like process
is identified in the optical. We find that the relatively high [OIII]
flux observed from the ground cannot be all produced in the nucleus.
Therefore, an extended NLR must exist in this object. This may be
interpreted in the framework of two different scenarios: i) the
radiation from ADAFs is sufficient to give rise to high ionization
emission-line regions through photoionization, or ii) the nuclear
source has recently ``turned-off'', switching from a high-efficiency
accretion regime to the present low-efficiency state.
\end{abstract}


\section{Introduction}
Low  luminosity active  galactic  nuclei (LLAGN)  are  believed to  be
powered  by accretion of  matter onto  the central  supermassive black
hole, similarly  to powerful  AGN. In a  large fraction of  LLAGN, the
central  black hole  is  as  massive as  in  powerful distant  quasars
($M_{BH}  \sim 10^{8}- 10^{9}  M_\sun$), thus  their very  low nuclear
luminosity  implies  that accretion  occurs  with  very low  radiative
efficiency      \citep[or     at      very     low      rates;     see
e.g.][]{ho04coz,papllagn}. If so, the physics of the accretion process
may be different from  the ``standard'' optically thick, geometrically
thin  accretion disks.   Starting from  the ``ion-supported  tori'' of
\citet{rees82}, a number of  theoretical models have been developed to
describe  such radiatively  inefficient  accretion flows  \citep[RIAF,
e.g.   advection-dominated accretion flows,  ADAF, advection-dominated
inflow-outflow solutions, ADIOS, convection-dominated accretion flows,
CDAF][]{narayanyi94,quataert99,abramowicz02}. But  because of the very
low radiation they emit at  all wavelengths, these objects (if they at
all  exist) are  very difficult  to  be observed.   Recently, the  AGN
nature of optical nuclear components seen in HST images of a sample of
LLAGN have been  unambiguously established.  \citet{Maoz05} have shown
that among  a sample of 17 LLAGN,  15 of them show  variability over a
timescale  of  a  few  months, which  demonstrate  their  non--stellar
origin. However, it  is still unclear whether the  radiation is from a
jet or from the accretion flow. LLAGN have also been found to lie
on  the  so-called  ``fundamental   plane  of  black  hole  activity''
\citep{merloni03,falcke04}, which attempts  to unify the emission from
all  sources around  black holes,  over a  large range  of  masses and
luminosities,  from  Galactic sources  to  powerful  quasars. But  the
origin of such  a ``fundamental plane'' and its  relationship with the
origin   of   the   radiation    is   still   a   matter   of   debate
\cite[e.g][]{bregman05,koerding06}.

RIAF  models  have  been  applied  to  several  sources  belonging  to
different classes,  such as low-luminosity  radio galaxies, ``normal''
ellipticals,       the       Galactic       center       Sagittarius~A
\citep[e.g.][]{quataert99,dimatteo00,dimatteo03}.   However,  for most
of these  objects, the models  cannot be properly  constrained, mostly
because the nuclear radiation is  swamped by other processes.  This is
particularly  problematic in  the optical  band, which  appears  to be
crucial to  fix the models, where  the stellar emission  from the host
galaxy is substantial.  It  is indeed in the IR-to-UV region that
different  accretion disk  models  are expected  to  show the  largest
difference  in  spectral  shape.   RIAFs  should lack  both  the  "big
blue-bump" and  the IR (reprocessed) bump,  which instead characterize
optically thick,  geometrically thin  accretion disk emission  and the
surrounding  heated  dust.   For  example, in  low  luminosity  radio
galaxies  non-thermal  emission from  the  jet  dominates the  optical
nuclear  radiation  \citep{pap1}, while  the  Galactic  center is  not
visible in the optical because it is hidden by a large amount of dust.

Therefore, neither of the above mentioned classes of objects appear to
be suitable  laboratories to  test models of  low-efficiency accretion
through  the analysis  of their  overall SED.   Recently,  among LLAGN
which  show very  low Eddington  ratios $L_{bol}/L_{Edd}  << 10^{-3}$,
\citet{papllagn} have found that a  class of LLAGN, mainly composed by
LINERs  and  low-luminosity  Seyfert~1  galaxies, show  faint  optical
unresolved  nuclei in  HST images  that may  be interpreted  as direct
radiation from  a very low  efficiency accretion flow.  In  fact, when
the radio-optical properties of  LLAGN are considered, Seyfert, LINERs
and low  luminosity radio galaxies separate into  different regions of
diagnostic planes,  according to the  properties of their  nuclei.  If
this  interpretation  is correct,  we  now  have  a powerful  tool  to
identify the  nature of the  nuclear radiation (i.e.   jet-dominated or
accretion-dominated).   The best  possibility  of detecting  radiation
from an ADAF-like  process would then be to  study unobscured Seyferts
of lowest luminosity, as well as  a sub-class of LINERs.  In all other
objects other radiation processes dominate.

 In  this   paper  we  further  test  the   picture  outlined  in
\citet{papllagn} by studying  the nuclear spectral energy distribution
of a galaxy,  \object{NGC~4565}, that seems to be  a perfect candidate
for hosting  a RIAF around  the central supermassive black  hole.  The
object is part of the ``Palomar sample'' of LLAGN \citep{ho97}, and it
is included in both the \citet{merloni03} and \citet{falcke04} samples
that  were  used to  define  the  ``fundamental  plane of  black  hole
activity''. It  is worth  mentioning that NGC~4565  does not  show any
significant peculiarity  in that plane.  NGC~4565 is  a nearby (d=9.7
Mpc)  LLAGN  classified  as  a  Seyfert~1.9 because  of  the  possible
presence  of a  faint,  relatively  broad (FWHM  =  1750 km  s$^{-1}$)
H$\alpha$ line.   However, as \citet{ho97broad} have  pointed out, the
detection of a broad component is highly uncertain.  As we show in the
following,  although it  is  a Type  2  Seyfert, this  object is  only
moderately  absorbed, and  the  nuclear radiation  is  visible in  the
optical spectral region.  NGC~4565  may thus represent the first clear
example of low-luminosity accretion  onto a supermassive black hole in
the optical band.

In  Section~\ref{obs}  we describe  the  {\it Chandra}  and {\it  HST}
observations, the  data analysis procedures and  flux measurements, in
Section~\ref{discussion}  we  present  the  results,  we  derive  the
spectral energy  distribution and  we discuss its  interpretation.  In
Section~\ref{conclusions}  we give  a summary of  our findings  and we
draw conclusions.

\begin{figure*}
\epsscale{1}
\plotone{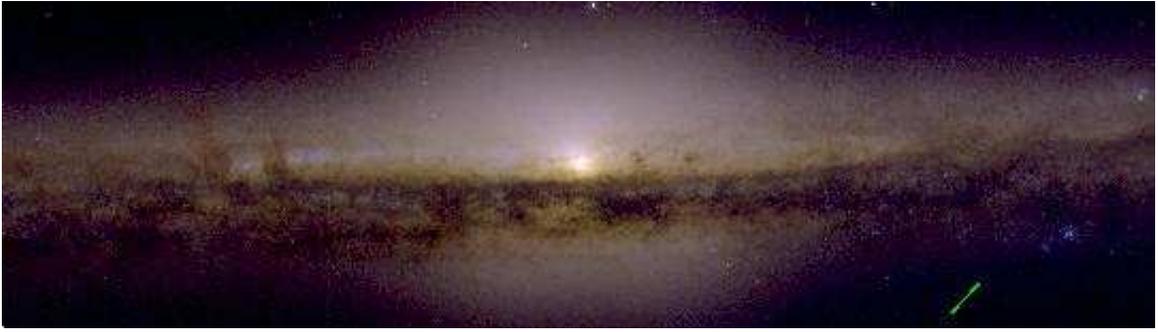}
\caption{HST WFPC2 ``true-color'' RGB image. The R-channel corresponds to the F814W filter, the B-channel is the F450W and the G-channel is the average of the two filters. Direction to the North is indicated by the arrow. \label{galaxy}}
\end{figure*}

\section{Observations and data analysis}
\label{obs}

We  use X-ray  data taken  with \facility{Chandra}  satellite,  and IR
through optical  \facility{HST} images.  In the following  we describe
the data and the analysis procedures.

\subsection{{\it Chandra} data}

A  60 ksec  ACIS-S  observation  of NGC~4565  (performed  in 2003,  PI
D.  Wang) is  publicly  available  in the  {\it  Chandra} archive.  We
retrieve the {\it  Chandra} data and analyze them  using standard CIAO
3.2.2 procedures,  applying the latest calibration files  in the CALDB
3.1.0 database. The X-ray image reveals a wealth of pointlike sources,
many  of which  located along  the  NGC~4565 disk.  The two  brightest
sources correspond to  an off-nuclear source at $\sim  50$ arcsec from
the nucleus, and  to the nucleus itself \citep[see  also the XMM image
in][]{foschini02}.

To avoid contamination from faint  nearby X-ray sources, the 0.4-7 keV
nuclear spectrum is extracted in a circular aperture of 6 pixel radius
($\sim  3$ arcsec, corresponding  to an  encircled energy  fraction of
$>97\%$ at  1.5 keV). The background  is evaluated in  a large annulus
around the nucleus.  Faint X-ray sources are not  masked out from the
background region, since their presence has a negligible impact on our
results  (the total background  flux including  faint sources  is less
than $1\%$ of the nuclear flux). Given the moderate nuclear count rate
(0.036 counts/sec),  X-ray photon  pileup is under  control ($\lesssim
4\%$).

Spectral analysis is  carried out with XSPEC v11.3.1,  with the column
density   of   our    Galaxy   fixed   to   $1.3\:10^{20}$   cm$^{-2}$
\citep{dickey90}. The spectrum is re-binned to have at least 20 photons
per bin to allow use of  the $\chi^2$ statistics, errors are quoted at
the  $90\%$ c.l.   for one  interesting  parameter.  We  find that  an
absorbed power law model provides  a very good description of the data
($\chi^2/dof=51/81$),  the best  fit photon  index and  column density
being   $\Gamma=1.91^{+0.22}_{-0.19}$   and   $N_H=2.5\pm0.6\;10^{21}$
cm$^{-2}$, respectively.  The observed  source fluxes in the 0.5-2 keV (soft)
and  2-10 keV (hard) bands  are $9.0\;10^{-14}$  erg  cm$^{-2}$ s$^{-1}$  and
$2.1\;10^{-13}$ erg cm$^{-2}$  s$^{-1}$, respectively.  When corrected
for absorption, these correspond  to intrinsic nuclear luminosities of
$1.9\;10^{39}$  erg s$^{-1}$  and $2.5\;10^{39}$  erg s$^{-1}$  in the
soft  and hard  band, respectively.   We note  that the  derived X-ray
spectral  parameters, fluxes  and luminosities  are in  good agreement
with those measured in a 14 ksec XMM-$Newton$ observation performed in
2001 \citep{cappi06}.

\subsection{{\it HST} data}

\begin{deluxetable}{l r c c c}
\tabletypesize{\scriptsize}
\tablewidth{0pt}
\tablecaption{Nuclear fluxes from HST observations \label{fluxes}}
\tablehead{\colhead{Instrument/Filter} & \colhead{T$_{exp}$} & \colhead{Program ID} & \colhead{Wavelength} & \colhead{$F_\lambda$} \\
                           \colhead{}  & \colhead{[s]}             & \colhead{}       & \colhead{[\AA]} & \colhead{ }\\}
\startdata
ACS-HRC/F330W  &  1200  &  9379 &  3367 & 8.1  \\
WFPC2/F450W    &   600  &  6092 &  4575 & 18   \\
WFPC2/F555W    &   320  &  6685 &  5468 & 21   \\
WFPC2/F814W    &   480  &  6092 &  8023 & 11   \\
NICMOS/F160W   &   384  &  7331 & 16074 & 5.9  \\
\enddata
\tablecomments{Fluxes (in units of $10^{-17}$ erg cm$^{-2}$s$^{-1}$  \AA$^{-1}$) have been corrected for local  extinction using $N_H=2.5\times
10^{21} cm^{-2}$ and standard $A_V/N_H = 5 \times 10^{-22}$ ratio.}
\end{deluxetable}

HST data are  available in the MAST archive at STScI  from the near IR
to  the  optical U-band.   Images  were  taken  as part  of  different
programs, with the following  instruments and filters: NICMOS (F160W),
WFPC2   (F814W,  F555W,   F450W),  ACS/HRC   (F330W).   These  filters
approximate  the H,I,V,B  and U  bands in  the HST    system. The
images  are  processed   with  the  standard  on-the-fly  reprocessing
calibration pipeline \citep[see][]{acshandbook}.

The optical images show the bulge of the galaxy partially covered by a
prominent  dust lane  or  disk seen  almost  edge-on (Fig. \ref{galaxy})
The inclination  of  the ``disk''  is  such that  the
central region of the bulge is  not covered by a large amount of dust,
and a  faint nuclear  compact source  (to which we  refer as  the {\it
nucleus}) is visible  in all images. 

\subsubsection{Nuclear photometry}

\begin{figure}
\epsscale{1}
\plotone{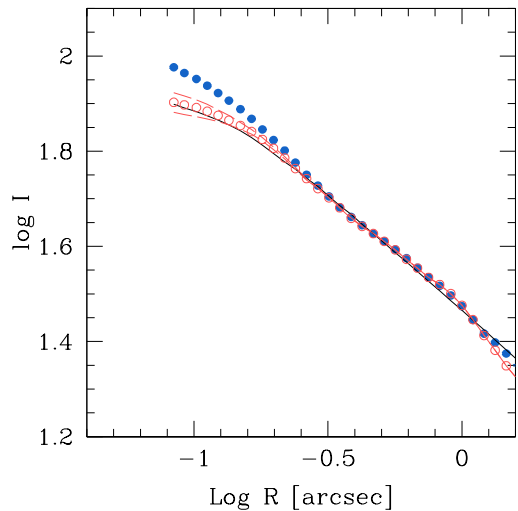}
\caption{Radial brightness profiles of  the central $\sim 1$ arcsec of
the  galaxy  used  to  measure  the flux  of  the  nuclear  unresolved
component. The filled circles are the observed profile (from the F814W
image),  the solid line  is the  model galaxy  convolved with  the HST
resolution.  The empty  circles are the derived profile  once that the
nuclear component has been subtracted  (see text for details). The two
dashed lines are obtained subtracting  PSFs that are 20\% brighter and
20\% fainter  than the flux  we report in  Table~1. The errors  on the
profiles  are  of  the order  of  or  smaller  than  the size  of  the
dots.\label{profiles}}
\end{figure}

In  the U-band,  the emission  from the  bulge stars  is low,  and the
nucleus is the by far the brightest source in the field of view of the
ACS/HRC  (Fig. \ref{profiles}).   Photometry  of the  nucleus is  thus
straightforward in  the U-band, also thanks to  the higher resolution,
smaller projected pixel-size of the HRC.  On the other hand, in the IR
(NICMOS) and optical I,V and B band WFPC2 images (the target is always
located in  one of  the WF cameras)  the contrast with  the underlying
stellar background is low, thus the measurement of the nuclear flux is
more problematic.   For the  photometry of nuclear  unresolved sources
superimposed to the stellar emission  of the host galaxy, we undertake
two different approaches, as described in the following.

1) Aperture  photometry  with the  {\it  IRAF  }  task {\it  radprof},
measuring  the  background  close  to  the unresolved  nucleus,  at  a
distance  of $\sim 0.4^{\prime\prime}$  from the  center of  the point
source, and  setting the  aperture radius at  the same  distance. Note
that the ``background'' here is  the stellar emission of the galaxy in
the vicinity of the nucleus.   Therefore, this approach works well for
nuclei in  elliptical galaxies  with flat radial  brightness profiles,
i.e.  Nuker-law  ``core'' galaxies  \citep{faber97} which have  a flat
($\gamma <  0.3$, $\Sigma \propto r^{-\gamma}$, where  $\Sigma$ is the
surface  brightness) slope  in the  inner region  \citep[see  also the
  discussion in][]{barbara}.   Clearly, this  is because in  this case
the ``background'' measured at a distance of $\sim 0.4^{\prime\prime}$
is  a good  estimate of  the  stellar emission  at the  center of  the
nucleus.  On  the other hand,  for both ``power-law''  ellipticals and
spirals bulges,  the profile  in the innermost  regions (i.e.   in the
central 1-2 arcsec) is  significantly steeper ($\gamma \sim 0.8$).  In
this latter case, the measurement and even the identification of faint
nuclei is  more difficult, because a ``peaked''  brightness profile of
the bulge  may hamper the detection  of the central  emission from the
AGN.   Furthermore, for  the IR  images,  which have  a lower  angular
resolution,  the background  cannot be  measured close  enough  to the
center of  the nucleus, and thus may  be significantly underestimated.
We find  that this would  lead to overestimate  the nuclear flux  by a
factor  as large  as $\sim  5$.  Thus,  while we  used this  method to
measure the nuclear flux in the F330W image \citep[aperture correction
  was   taken  into   account,  following   the   prescriptions  given
  in][]{sirianniacs},  we had to  adopt a  different strategy  for the
WFPC2 and NICMOS images.

2) An alternative approach consists  in deriving the radial brightness
profile  of the  galaxy,  and measure  any  nuclear excess.   Multiple
component models  are often used  to reproduce the central  regions of
galaxies and measure the flux  of nuclear sources \citep[see e.g.  the
  discussion in][]{quillen01}.   But since here we  are not interested
in  modeling the  galaxy bulge  on large  scales, we  only  derive the
radial brightness  profile in  the central $\sim  2$ arcsec.   Then we
produce a model  galaxy with the same slope as  observed in the region
$R > 0.2$ arcsec and we assume that the profile can be extrapolated to
the  center of the  bulge, all  the way  to $R=0$.   As shown  in Fig.
\ref{profiles} (solid  line), the effect  of the finite  resolution of
HST ($\sim 0.1$ arcsec) produces a flattening of the observed profile,
at a distance  of $\sim 0.15$ arcsec.  The  observed profile, obtained
by fitting ellipses to the galaxy image using the {\it IRAF} task {\it
  ellipse}, shows  a significant excess (filled  circles).  We produce
synthetic PSF's using the {\it TinyTim} software \citep{krist}, which,
for WFCPC2, produces an  accurate representation of the central region
of the PSF, thus appropriate  for our purpose.  We align the synthetic
PSF's with the  position of the nucleus, we multiply  the PSF image by
an appropriate constant $K$ and we subtract the two images.  We change
the  value of  $K$ until  the profile  of the  nuclear regions  do not
produce a ``hole'' at the  center of the galaxy.  The obtained profile
is shown in Fig.  \ref{profiles} as the empty circles.  To convert the
flux of  the nucleus  from counts to  physical units, we  multiply the
count rate ($CR = K/t_{\rm exp}$) by the keyword {\it PHOTFLAM} in the
image header (for NICMOS  $CR=K$). In Fig.~\ref{profiles} we also show
the  radial  profiles  obtained  by  subtracting PSFs  that  are  20\%
brighter and 20\% fainter than our reference value. 

It  is  not  straightforward  to   estimate  the  error  on  the  flux
measurements obtained using method  2, because the main uncertainty is
the  assumption  that  the  radial   profile  of  the  galaxy  can  be
extrapolated all  the way  to the center,  at R=0.   After subtracting
synthetic PSFs with different total counts and comparing the resulting
profiles  with  our  model  profile,  we  prefer  to  adopt  a  rather
conservative value of $\sim 20\%$ for  the error of the IR and optical
fluxes.   With future  observations,  which should  be  taken using  a
dithering  strategy aimed  at improving  the PSF  sampling,  the error
could be  significantly reduced.  The  statistical error on  the F330W
flux (obtained with method 1) is  7\%.  A summary of the photometry is
given in Table~\ref{fluxes}.

The  F450W  and  F555W  filter  pass bands  include  relatively  strong
emission lines  (mainly [OIII]5007 and H$\beta$).   However, since the
pass bands are $\sim 1000$\AA~~ wide, the observed flux is likely to be
dominated by continuum emission (see also Sect. \ref{discussion}).

Since the near-IR and the  U-band images were not taken simultaneously
to  the optical  data, the  SED may  be affected  by  variability.  We
checked for  variability of the  nuclear source in the  optical (F814W
and F450W), for which two  sets of observations with the same filters,
taken  at a distance  of $\sim  1$ year,  are available.   The nuclear
fluxes are consistent within the  errors, thus no variability is found
between the  2 observations.  However,  our estimated $20\%$  error on
the optical fluxes does not  allow us to exclude variations of smaller
amplitude, as observed in other LLAGN \citep{Maoz05}.

\begin{figure}
\epsscale{1}
\plotone{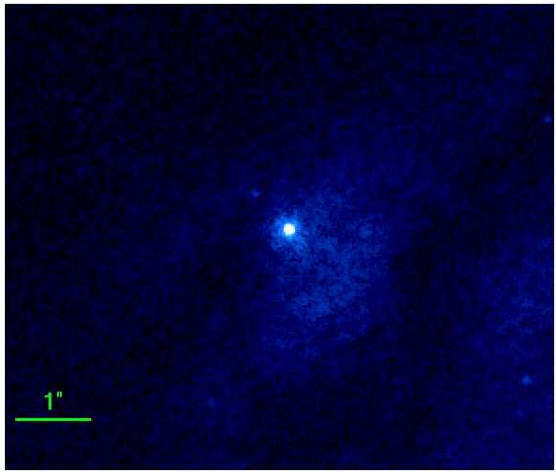}
\caption{HST ACS/HRC image in the  U-band (F330W filter). North is up,
East is left (left panel).   \label{uband}}
\end{figure}

\section{Results and discussion}
\label{discussion}

\begin{figure}
\plotone{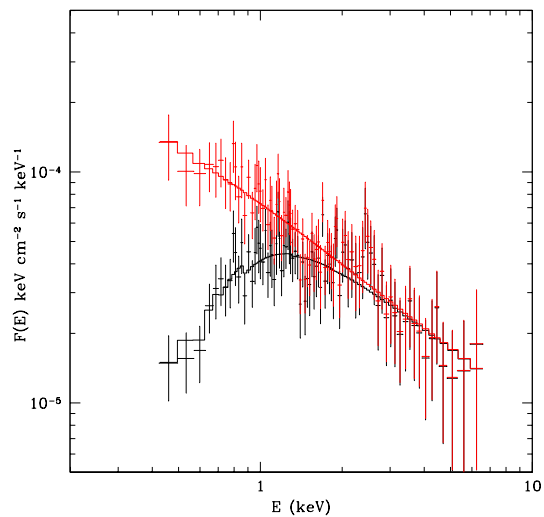}
\caption{X-ray spectrum  from Chandra data. The  lower data-points and
solid line correspond  to the absorbed data and  model outlined in
text.  The  upper  data-points   and  solid  line  correspond  to  the
unabsorbed data and model.  \label{xspec}}
\end{figure}

\begin{figure}
\epsscale{1}
\plotone{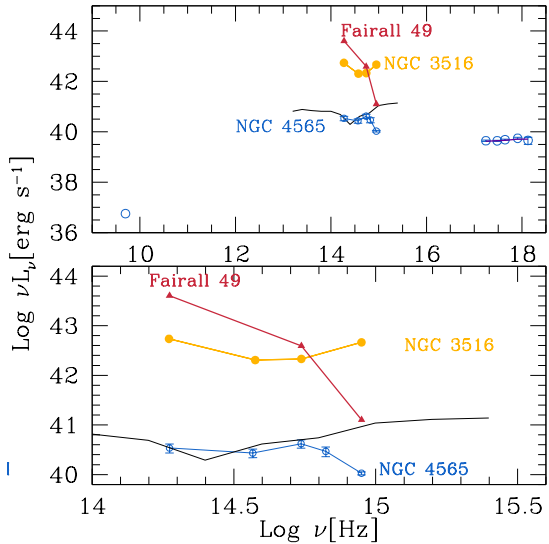}
\caption{Absorption corrected nuclear  spectral energy distribution of
NGC~4565 from  the radio  to the X-ray  band.  The X-ray  spectrum has
been  significantly re-binned to  improve the  clarity of  the figure.
The solid  line superimposed to the  X-ray data is  the spectral model
used to fit the data (see text).  For comparison, we show the IR-to-UV
SED  of  a  Seyfert~1  (NGC~3516)  and  of  a  Compton-thin  Seyfert~2
(Fairall~49). The  solid line  is the average  SED of  radio-quiet QSO
from \citet{elvis94}, normalized to the  flux of NGC~4565 in the F814W
filter.   The  lower  panel  is  a zoom  into  the  IR-to-UV  spectral
region. \label{sed}}
\end{figure}

\subsection{The nuclear SED of NGC~4565}

The absorption corrected nuclear spectral energy distribution is shown
in  Fig.~\ref{sed}.  The  HST data  are de-reddened  using $N_H  = 2.5
\times 10^{21}$  cm$^{-2}$, which converts  to $A_V =  1.25$, assuming
Galactic gas-to-dust ratio.  Although in AGN the gas-to-dust ratio may
differ from the local value,  we believe that this choice is justified
in the case of NGC~4565.  As it  is clear from the large field of view
image of the galaxy (Fig.~\ref{galaxy}),  this is a spiral seen almost
edge on.   Therefore, it  is reasonable to  assume that  a significant
amount  of dust  and gas  in the  disk of  the galaxy  project  on our
line-of-sight to the nucleus.

Assuming  a circular geometry,  from the  observed ellipticity  of the
disk in the  image we find that the orientation of  the disk is likely
not  to exceed  $\sim 10^\circ$.   For  comparison, we  can check  the
absorption   we  find   in  our   Galaxy  for   $10^{\circ}$  Galactic
Latitude. We  obtain $A_V \sim  1.0 - 1.5$,  where the lower  value is
found for Galactic longitude $\sim 180^\circ$, the higher value is for
$\sim  0^\circ$  (from  NED).   These  values  may  actually  increase
substantially if we observe the same Galactic Latitude from the Galaxy
center.  This simple check shows that the absorption we measure in the
X-rays is compatible with that  provided by galactic dust in the disk.
This supports our hypothesis  that the moderate absorption observed to
the nucleus  of NGC~4565 is not  produced locally, in  the vicinity of
the nucleus. In  this case, Galactic dust-to-gas ratio  may be used to
convert $N_H$ derived from the X-rays to optical $A_V$.

The  nuclear  SED appears  basically  flat  ($\alpha  \sim 1$,  $F_\nu
\propto \nu^{-\alpha}$) from the  1.6$\mu$m to 4500 \AA, with possibly
a small peak  between 5000 and 4000 \AA~~ and a  small drop-off in the
U-band.  This  peak may  be real, or  due to a  possible contamination
from  emission lines  (mainly [OIII]  and H$\beta$)  that fall  in the
F555W  and F450W  filters  pass bands.   Unfortunately, since  neither
images with  narrow-band filter nor  nuclear spectra are  available to
date, a  certain ambiguity persists.   However, all other  filters are
free from strong lines, thus  the intrinsic SED cannot be dramatically
different from what we show here.  Whatever the nature of such a small
peak, it  is clear that neither  a significant UV bump  nor IR thermal
emission from hot dust, which  are characteristic of AGNs, are visible
in NGC~4565. Furthermore, note that the luminosity in the X-ray is not
higher than in the optical,  even after absorption has been taken into
account (see Fig.~\ref{xspec}).

For  comparison, in  Fig.~\ref{sed} we  show  the nuclear  SED of  two
specific objects, together with the  average SED of radio-quiet QSO as
taken  from \citet{elvis94}. The  two objects  are a  Seyfert~1 galaxy
(\object{NGC~3516})  and  of  a Seyfert~2  (\object{Fairall~49}),  for
which the absorption, estimated from X-ray observations, is $N_H = 1.4
\times  10^{22}$ \citep{iwasawa04}.  These  two Seyferts  have similar
bolometric luminosity,  but they are  both clearly more  powerful than
NGC~4565, by $\sim 3$ orders  of magnitude.  The Type~1 object clearly
shows a concave spectrum, which is interpreted as the signature of the
presence of the blue bump in the  UV and of dust heated by the central
AGN in the near-IR.  The Type~2 galaxy, instead, is very bright in the
IR,  while  the flux  is  dramatically  reduced  for higher  (optical)
frequencies,  as  a   result  of  absorption.\footnote{It  might  seem
surprising to detect  the nucleus in Fairall~49, which  is a Seyfert~2
absorbed  by a  large amount  of  $N_H$ in  the X-rays.   In fact,  if
converted  to $A_V$  using standard  Galactic dust-to-gas  ratio, this
would correspond  to 7 mag extinction in  V and 11.5 mag  in the F330W
filter. Two possible  explanations have been proposed: i)  part of the
optical nuclear flux in Fairall~49, if not all, might not be radiation
from the  accretion disk seen  directly (through a moderate  amount of
dust).  Instead, the nucleus might be in part (or completely) obscured
in the optical-to-UV  band, and the bulk of  the observed emission may
be scattered  light; ii) the properties of  the circumnuclear absorber
are  different from  the  Galactic dust  and  this would  result in  a
non-standard  $A_V/N_H$ ratio.   This  has been  discussed by  various
authors \citep[e.g.][]{granato97,maiolino01}.}   Not all Seyferts show
such a clear behavior, but these  objects serve as good examples to be
compared with the peculiar SED of NGC~4565.

\begin{figure}
\epsscale{1}
\plotone{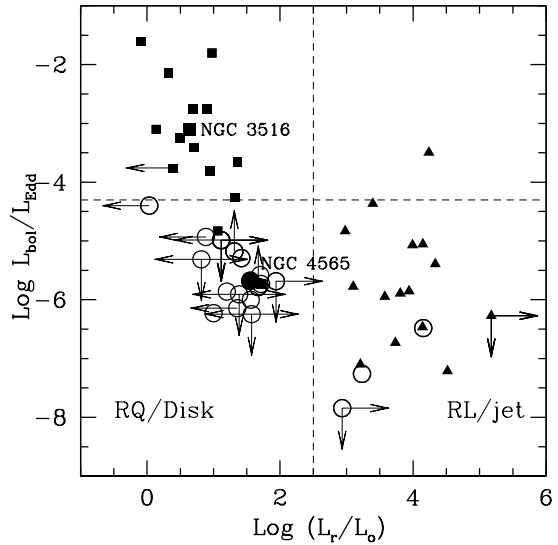}
\caption{Optical  to Eddington  luminosity ratio  plotted  against the
radio to optical ratio (the ``nuclear radio-loudness'') for the sample
of  nearby  LLAGN  \citep[adapted from][]{papllagn}.   Seyfert~1s  are
plotted  as  squares, low  luminosity  radio  galaxies are  triangles,
LINERs  are  empty circles.   NGC~4565  (filled  circle) and  NGC~3516
(large square)  are also marked in  the figure.  The  dashed lines are
only  used  to guide  the  eye  and divide  objects  of  high and  low
Eddington  ratio  (top  and  bottom  of the  figure)  and  radio-quiet
(disk-dominated  nuclei, left)  or radio  loud (jet  dominated nuclei,
right). \label{r_edd}}
\end{figure}

\subsection{A low radiative efficiency accretion disk}

How do we interpret  the nuclear emission in NGC~4565? As pointed
out in the introduction, the source  is included in the sample of both
\citet{merloni03} and \citet{falcke04}. The  object does not appear to
show any  peculiarity, and its location along  the ``fundamental plane
of black hole  activity'' does not provide us  with information on the
emission process.   Therefore, in order to answer  the above question,
we  use  diagnostics  introduced  by  \citet{papllagn}.   As  already
mentioned in  Section~1, we  found that the  radio to  optical nuclear
luminosity ratio (i.e.   the ``nuclear radio-loudness'') $L_r/L_o$
for LLAGNs gives us important information on the nature of the source.
In  particular, we  can  infer whether  we  are observing  synchrotron
emission  in both  bands (if  $\log (L_r/L_o)\sim  3$), or  in the
optical we  have some kind of  excess radiation which, in  the case of
unabsorbed Seyferts  is most likely interpreted as  radiation from the
accretion  process.   Furthermore,  when  the  optical  luminosity  to
Eddington  luminosity  ratio  $L_o/L_{Edd}$  is  plotted  against  the
nuclear radio loudness (Fig.  \ref{r_edd}) different classes of LLAGNs
nicely separate into three  different regions of the diagram.  Seyfert
nuclei with  relatively high  accretion efficiency objects  occupy the
top-left part  of the diagram ($L_o/L_{Edd}  \sim 10^{-2}-10^{-3}$ and
$\log(L_r/L_o)  \sim  1$); LINERs  separate  into two  subclasses,
which  we named  according  to their  nuclear  radio-optical ratio  as
``radio--quiet'' LINERs (bottom-left side, $L_o/L_{Edd} < 10^{-4}$ and
$\log(L_r/L_o)    \sim    1$),    and    ``radio--loud''    LINERS
(bottom-right); radio galaxies (bottom-right part of the diagram) have
the same Eddington ratio as for radio--quiet LINERs, but a much higher
$\log(L_r/L_o)$. Note that in  the plane of Fig.  \ref{r_edd}, the
objects  in which  we observe  an extra-component  in the  optical, in
excess of synchrotron  emission, are those that lie  on the left side.
Therefore, those are the objects  in which emission from the accretion
process can be detected.    In fact, in perfect agreement with the
shape of its SED, NGC~3516 lies  in top-left panel of the plane, where
relatively high efficiency accretion disks are.  On the other hand, we
did not  mark the location of  the Seyfert~2 galaxy  Fairall~49 on the
plot of Fig.~\ref{r_edd}, since the nuclear emission is most likely to
be  affected by significant  obscuration. However,  we  point out
that assuming that the IR flux  can be used as instead of the optical,
the radio to IR  ratio would be similar to the Seyfert  1s in the plot
(at $\log L_r/L_o = 0.88$).

Let  us  explore how  this  applies to  NGC~4565.   First  of all,  we
calculate the  ratio between  the nuclear radio  flux and  the optical
flux.  \citet{neil05}  measured a  radio core flux  of 3.2  mJy, which
implies that  $\log(L_r/L_o) =  1.6$.  Therefore, NGC~4565  has an
excess in the  optical of at least 2 dex with  respect to the expected
synchrotron emission (i.e.  the  optical counterpart of the radio core
should be $> 2$ dex fainter than the measured optical flux, unless the
radio-to   optical  spectral   index  has   unreasonable   values  for
synchrotron  radiation).   Thus  it  is reasonable  to  interpret  the
optical  nucleus as  radiation from  the accretion  process.   In this
case, assuming a central black hole mass of $2.8 \times 10^{7} M_\sun$,
as  derived from  the  M-$\sigma$ relation  of \citet{tremaine02},  and
using        $\sigma$         value        from        the        LEDA
database\footnote{http://leda.univ-lyon1.fr/}, the resulting Eddington
ratio  $L_o/L_{Edd}$ is  extremely  low, $2  \times  10^{-6}$.  It  is
important to note that in the case of ``typical'' Seyferts, which show
a blue  bump, a significant bolometric correction  is needed (however,
this should  not exceed a  factor of $\sim  15$).  For NGC~4565  a big
blue bump  is clearly not  present, thus our  value of $L_o$  is a
good   estimate  of   $L_{bol}$.    

In the  diagnostic diagram of  Fig.~\ref{r_edd}, NGC~4565 lies  in the
lower-right  quadrant, among  ``radio-quiet'' LINERs  \citep[the other
two Seyferts in the same region  of the plot are M~81 and NGC~4639, as
discussed in][]{papllagn}.  In order  to reconcile NGC~4565 with other
Seyferts,  which are confined  in the  top-left quadrant,  the central
black  hole mass would  have to  be at  least a  factor of  $\sim 100$
lower.   This  would  substantially  violate  the  $M_{BH}  -  \sigma$
relation.   We  conclude  that  the  nucleus of  NGC~4565  is  a  very
low-efficiency  accretion  object  and   that  we  are  observing  the
accretion  process  directly  in   the  optical.   This  is  extremely
important  since  models of  advection-dominated  accretion flows  are
particularly  sensitive  to   the  optical-UV  spectral  region.   For
example, as shown  by \citet{quataert99} the presence of  winds in the
disk  can dramatically change  the shape  of the  observed SED  in the
range  of frequencies  between $\nu  \sim  10^{13}$ Hz  and $\nu  \sim
10^{15}$ Hz.   However, a detailed comparison with a set of models
is beyond the scope of this  paper.  RIAF models are very sensitive to
many  different physical parameters,  therefore detailed  modeling is
useful only  if the SED is  well determined, from the  radio-mm to the
X-ray band and, if possible, when simultaneous data are available.

A  similar  study  of  the  nuclear emission  has  been  performed  by
\citet{moran99}  for NGC~4395,  ``the least  luminous  Seyfert~1''. In
that case, the nuclear luminosity  is even lower than in NGC~4565, but
the  central black  hole mass  in NGC~4395  is dramatically  lower.  A
recent  estimate  based on  reverberation  mapping  gives  a value  of
$M_{BH} =  3.6 \times 10^{5}  M_\sun$ \citep{peterson05}.  Such  a low
black hole  mass implies $L_{bol}/L_{Edd} \sim 10^{-3}$  or higher if,
as \citet{moran99} point out, intrinsic nuclear absorption is present.
This seems  in fact to be  the case since  \citep{moran05} obtain $N_H
\sim 10^{22}$ cm$^{-2}$ analyzing {\it Chandra} data, although most of
the absorption  might be produced by ionized  gas. Therefore, although
it is  clear that even if NGC~4395  displays peculiar characteristics,
its  Eddington  ratio  is  not  different  from  the  average  of  low
luminosity    Seyferts    in    the    Palomar   and    CfA    samples
\citep{papllagn,ho04coz}.  Instead,  NGC~4565 has completely different
physical properties,  as appears from  its extremely low value  of the
Eddington ratio, and it is a perfect candidate for hosting a radiative
inefficient accretion process.

\subsection{Where is the narrow line region in NGC~4565?}

One further implication  of the observations we present  here is worth
mentioning.  The  flux of the [OIII]5007 emission  line ($F_{[OIII]} =
1.5  \times 10^{-14}$ erg  s$^{-1}$ cm$^{-2}$),  as measured  from the
ground  with a  2'' beam  size \citep{ho97},  and the  diagnostic line
ratios  are typical  of Seyfert  galaxies.   Even  considering the
interesting (although  only slightly different)  classification scheme
for LLAGN  proposed by \citet{kewley06}  based on SDSS  data, NGC~4565
still  falls in  the region  occupied  by Seyferts.   However, if  we
assume  that all  of the  [OIII] flux  is produced  in  the unresolved
nucleus, this would  result in a count rate higher by  factor of 5 and
20  than  we   measure  in  the  nucleus  in   the  F555W  and  F450W,
respectively.  This implies that the narrow emission line region (NLR)
must be  extended.  It is particularly interesting  to investigate the
properties of the  NLR relatively to the nuclear  properties, since it
is  sometimes  assumed  that  radiative inefficient  accretion  cannot
provide a sufficient photon field to ionize the surrounding medium and
create the NLR.   If this is true, we can  speculate that NGC~4565 may
have recently transitioned from a relatively high-efficiency accretion
state (as  in ``normal'' Seyferts) to a  very low-efficiency accretion
process.  This might also reconcile its classification as Seyfert with
the fact  that its  nucleus is  located among LINERs  in the  plane of
Fig.~\ref{r_edd}.  The spectral classification is in fact based on the
large-scale  properties of the  emission-line gas,  that may  still be
powered by a higher radiation  field (possibly having also a different
spectral shape),  because of light travel time  effects.  However, the
equivalent width of  the [OIII] line, as measured  with respect to the
nuclear continuum  emission, is $EW_{[OIII]} \sim  100$\AA. This value
is   in  the  range   normally  spanned   by  Type~1   AGN  \citep[see
e.g.][]{kinney91,pap4,marziani03}.  Therefore,  this may indicate that
the continuum  emission from  RIAFs is sufficient  to account  for the
observed [OIII] flux, and Seyferts'  NLR can be powered by radiatively
inefficient accretion  flows.  However, in  the scenario in  which the
accretion disk has  changed its state, $EW_{[OIII]}$ may  assume a low
value if only a fraction of  the NLR gas is still highly ionized under
the effect  of the ``past''  high--efficiency state of  the accretion.
Clearly, only high spatial-resolution images with narrow band filters,
and nuclear spectra, can provide further information to understand the
recent history and  ionizing mechanism of both the  nucleus and NLR of
NGC~4565.

\section{Summary and conclusions}
\label{conclusions}

We  have  derived  the  spectral  energy distribution  of  a  peculiar
low-luminosity   Seyfert~2   galaxy   which,  despite   its   spectral
classification,  basically   shows  no  evidence   for  local  nuclear
absorption. The SED is peculiar, as it is almost flat in a $\log \nu -
log (\nu  F\nu)$ representation, with  no sign of  both a UV  bump and
thermally  reprocessed IR emission.   The very  low luminosity  of the
source associated with a relatively high central black hole mass imply
an  extremely small value  of the  Eddington ratio  ($L_o/L_{Edd} \sim
10^{-6}$). This, together with the position occupied by this object on
diagnostic planes  for low  luminosity AGN, represents  clear evidence
for  a low  radiative  efficiency  accretion process  at  work in  the
innermost  regions  of  NGC~4565.   The direct  detection  of  optical
emission  from such  radiative inefficient  processes  is particularly
important  for  providing  constraints  to  ADAF  models  or  similar.
NGC~4565 is therefore a perfect candidate for studying RIAFs, and more
observations aimed at  achieving a complete coverage of  the SED, from
the radio-mm to the X-ray bands,  should be taken in order to test the
models.  As part of a ``search for RIAFs in LLAGN'', it would also
be extremely useful  to derive the SED of objects  that are located in
the same region of the diagnostic planes as NGC~4565.

The fact  that the  [OIII] emission line  flux is substantial  in this
object implies that  an extended narrow line region,  similar to other
Seyfert galaxies, is still  present in NGC~4565. A possible intriguing
scenario  is  that the  active  nucleus  has recently  ``turned-off'',
switching  from a  high  efficiency, standard,  accretion  disk, to  a
radiative inefficient accretion process.  However, since the EW of the
[OIII]5007 emission  line is  rather small, with  the present  data we
cannot rule out  that the amount of ionizing photons  from the RIAF is
sufficient to produce the observed [OIII] flux.

\acknowledgments

We  thank the  anonymous  referee for  her/his  comments that  greatly
improved the  paper.  We acknowledge Dave Axon,  Alessandro Capetti and
Alice Quillen  for  useful comments.  RG
acknowledges support from the STScI Visitor Program.

This  research has made  use of  the NASA/IPAC  Extragalactic Database
(NED) which  is operated by the Jet  Propulsion Laboratory, California
Institute of Technology, under  contract with the National Aeronautics
and Space Administration.



{\it Facilities:}  \facility{HST}, \facility{Chandra}.



\end{document}